\documentclass[aps,prm,showpacs,preprint,groupedaddress]{revtex4-1}
\usepackage{epsfig}
\usepackage{dcolumn}
\usepackage{epstopdf}
\usepackage{graphicx}
\usepackage{ulem}
\usepackage{color}
\begin{document}
\title{High thermoelectric figure of merit and thermopower in  layered perovskite oxides}
\author{Vincenzo Fiorentini}
\author{Roberta Farris}
\author{Edoardo Argiolas}
\author{Maria Barbara Maccioni}
\affiliation{Department of  Physics at University of Cagliari, and CNR-IOM, UOS Cagliari, Cittadella Universitaria, I-09042 Monserrato (CA), Italy}
\date{\today}

\begin{abstract}
We predict high thermoelectric efficiency in the layered perovskite La$_2$Ti$_2$O$_7$, based on calculations (mostly ab-initio) of the electronic structure, transport coefficients, and thermal conductivity in a wide temperature range.  The figure of merit $ZT$ computed with a temperature-dependent relaxation time increases monotonically from just above 1 at room temperature to over 2.5 at 1200 K, at an optimal carrier density of around 10$^{20}$ cm$^{-3}$. The Seebeck thermopower coefficient is between 200 and 300 $\mu$V/K at optimal doping, but can reach nearly 1 mV/K at low doping. Much of the potential of this material is due to its  lattice thermal conductivity of order 1 W/(K m);  using a model based on ab initio anharmonic calculations,  we interpret this low value as due to effective phonon confinement within the   layered-structure blocks.
\end{abstract}
\maketitle

\section{Introduction}
The interest in thermoelectricity as an energy source increases steadily, and many old and new materials are now being assessed as candidate thermoelectrics  via  their figure of merit 
\begin{equation} {ZT}=\frac{\ \sigma S^2}{\kappa} T,
\end{equation}
with $\sigma$ and $\kappa$ the electrical and thermal conductivities, $S$ the Seebeck coefficient, and $T$ the temperature.  Recently, we have analyzed theoretically Mg$_3$Sb$_2$ \cite{noi2,noi1}, a material that is by now relatively mainstream in basic thermoelectricity research \cite{mgsb1}, with some success. This encourages us to venture into assessing the potential of more unusual materials.
In this paper we study  La$_2$Ti$_2$O$_7$ (henceforth LTO), the standard-bearer of the family of layered perovskite oxides 
A$_n$$X_n$O$_{3n+2}$. Several of these materials, are  high-T$_c$ ferroelectrics \cite{anewFE,iniguez,lichtenb}, and have  potential for other exotic effects such multiferroicity \cite{noi-vlto,noi-phtr} and metallic ferroelectricity  \cite{noi-BiTO}. LTO, in particular, is a wide-gap insulator, and a ferroelectric with Curie temperature of 1770 K, and with polarization parallel to the $c$ axis. 

\begin{figure}[ht]
\centerline{\includegraphics[width=0.6\linewidth]{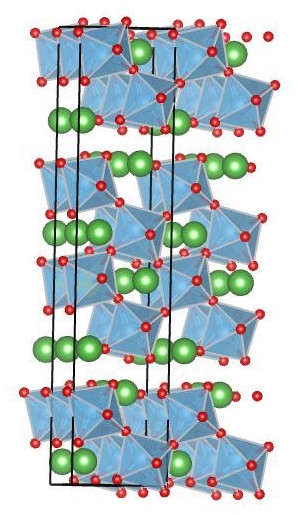}}
\caption{\label{struc} Sketch of the structure of LTO. The $a$ axis points into the page, the $b$ axis vertically, and the $c$ axis from left to right. The primitive cell is outlined.}
\end{figure}

While $ZT$ is at most 0.3 at high $T$ \cite{peng} in standard doped perovskites, the layered variant LTO  possesses  features  suggesting  promise for thermoelectric efficiency. First, the conduction band has  a rapidly rising density of states (of $s$-$d$ Ti character in the conduction band and mostly O $p$ in the valence) which according to the Mott-Cutler formula \cite{mott} should produce a significant Seebeck coefficient. Second, the layered structure of perovskite blocks (consisting of four perovskite octahedral units  stacked along a (110) direction in cubic perovskite axes,  as sketched in Figure \ref{struc}) may be expected to hinder phonon propagation, much like  a multi-interface structure, possibly leading to a reduced lattice thermal conductivity compared to normal perovskites. Experimental indications, though limited, support these views: in Ref.\cite{expseeb} an encouraging Seebeck coefficient of 120 to 300 $\mu$V/K in the 200 to 800 K range was measured in a sintered-ceramic sample; and Ref.\cite{klto} reported an almost temperature independent lattice thermal conductivity of about 1.2 W/(K m) in nominally undoped, crystalline LTO. 

The goal  of this paper is the assessment of transport coefficients and thermoelectric figure of merit in LTO. Subordinately, we aim at  qualitatively explanaining  its small lattice thermal conductivity.  For the electronic coefficients (electrical and electronic thermal conductivity, Seebeck thermopower coefficient) we calculate the electronic band structure from first principles within density functional theory in the generalized gradient approximation, and then use it to calculate the coefficients as function of temperature and doping, in  the relaxation-time approximation to the linearized  Boltzmann transport equation known as Bloch-Boltzmann theory \cite{allen,bt2}. We then use a $T$-dependent relaxation time to obtain $ZT$. 
The lattice thermal conductivity is  estimated with   a  Boltzmann equation for phonons \cite{tk} which include { {\it ab initio}} third-order anharmonic scattering  \cite{D3Q}, isotopic scattering, and Casimir finite-size boundary scattering.
We concentrate on $n$-type doping, and compare it with less efficient $p$-doping later in the paper. In our  $T$ range, which we choose to be  300 to 1200 K to avoid getting too close to the ferroelectric transition at 1800 K, the Seebeck coefficient is  between about 200 and 300 $\mu$V/K, and the  lattice thermal conductivity is only about 1 W/(K m) and almost $T$-independent. As a result, we find that the figure of merit increases monotonically from over 1 at room temperature to over 2.5 at 1200 K   within a $T$-dependent scattering-time approximation (on the other hand, in the constant-time approximation, $ZT$  exceeds 5.5 at the upper end of the range).

\section{Methods}

{
 The methods in this paper combine ab initio calculations with phenomenological and model approaches. As will be described below, electronic bands are treated ab initio, but transport coefficients are obtained in the framework of semiclassical Boltzmann theory, and electron scattering times are calculated via model expressions and thermal averaging; similarly, the lattice thermal conductivity is estimated via a model  based  on a Boltzmann equation for phonon transport including, among others, ab initio anharmonic phonon-phonon scattering terms. Overall the various approximations are internally consistent, and adopted conservatively in all cases.}

\subsection{Electronic structure and  transport coefficients}

{ Ab initio calculations are performed within the generalized gradient approximation \cite{pbe} using the VASP code \cite{vasp} and the projector augmented wave method, with the maximum suggested cutoff and the {\tt La}, {\tt \verb+Ti_pv+}, and {\tt \verb+O_s+} PAW datasets. The  structure of the orthorombic phase of LTO with space group $Cmc2_1$ is optimized following quantum forces and stress. The computed lattice constants are $a$=3.915 \AA, $b$=25.851 \AA, $c$=5.643 \AA; the internal positions are provided in the Supplementary Material. The electronic eigenvalues are also calculated ab initio on a (24$\times$8$\times$16) k-points grid. The conduction band minimum is at 
$\Gamma$, while the valence band maximum is at $X$=($\pi$/$a$,0,0); the minimum gap in GGA is 2.85 eV and becomes about 4.1 eV applying the correction of Ref.\cite{fb}.}
   The  BoltzTrap2 (BT2) \cite{bt2} code is then used for the calculation of the coefficients. The ab initio bands (assumed   rigid, i.e. not changing with doping or temperature) are interpolated   by a Fourier-Wannier technique \cite{bt2} over  a number of k-points given by the original number of points times an amplification factor, which we choose to be $A$=64, which is approximately equivalent to a much finer  (96$\times$32$\times$64) grid. 

\subsection{T-dependent relaxation time}

$ZT$ must be obtained from the various coefficients under some hypothesis for the relaxation time. 
Indeed if (as  discussed in Ref.\cite{noi2}) the relaxation time is assumed to be a constant 
$\tau_0$, it will factor out of all the integrals determining the Onsager coefficients; the 
BT2 code will then return the coefficients $\overline{\sigma}_0$=$\sigma$/$\tau_0$ and $
\overline{\kappa}_{e,0}$=$\kappa_e$/$\tau_0$,  determined  by the band-
structure and by temperature, but independent of $\tau_0$. Since the lattice thermal 
conductivity is non-zero, an estimate of $\tau_0$ must be provided to calculate $ZT$ in the 
constant-time approximation,
 \begin{equation}
 { ZT}_0=\frac{\sigma S^2}{\kappa_{e}+\kappa_{\ell}}=\frac{\overline{\sigma}_0\tau_0 S^2}{\overline{\kappa}_{e,0}\tau_0+\kappa_{\ell}}.
 \end{equation}
This  depends strongly on $\tau_0$  precisely in the region of typical expected values (see e.g. Figure 1 of Ref.\cite{noi2}).  Aside from 
 uncertainties in determining it, a fixed $\tau_0$  neglects the physically relevant temperature dependences due to phonon occupation changes and energy averaging. To obviate this 
 shortcoming, we  calculate 
\begin{equation}{ ZT}=\frac{\overline{\sigma}_0\tau_{\rm ave}(T) S^2}{\overline{\kappa}_{e,0}\tau_{\rm ave}(T)+\kappa_{\ell}}
\label{ZTT}
\end{equation}
using a  T-dependent scattering time, which is chosen to be the energy-average  \begin{equation}
\tau_{\rm ave}(T)=
\frac{\int_0^{\infty} \tau(T,E) E^{3/2} (-\frac{\partial f_{\rm F}(T, E)}{\partial E})\, dE}{\int_0^{\infty}
 E^{1/2} f_{\rm F}(T, E)\, dE}
\label{tauave}
\end{equation}
for a parabolic band (Ref.\cite{cardona}, Eq.5.21) of a $T$- and energy-dependent relaxation time. Here  $f_{\rm F}$ is the Fermi-Dirac distribution, and  the  relaxation time is
\begin{equation}
\tau(T,E)=\frac{1}{P_{\rm imp}+P_{\rm ac}+P_{\rm polar}}.
\label{scrat}
\end{equation}
 obtained from the rates of impurity, acoustic phonon and polar phonon scattering.
 The model expressions of the $P$'s in Eq.\ref{scrat} are given in Refs.\cite{noi2,ridley1,ridley2}.  
  \begin{figure}[ht]
\centerline{\includegraphics[width=0.8\linewidth]{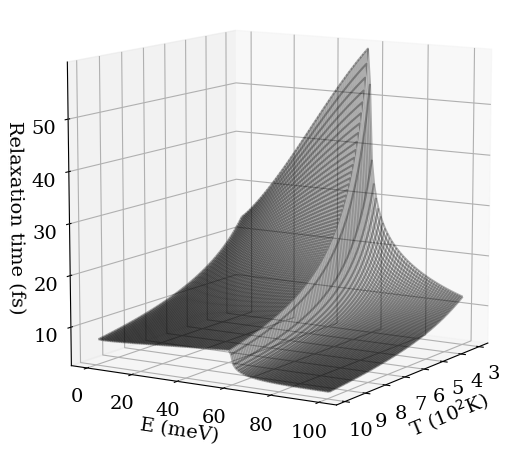}}
\caption{\label{taufig1} Relaxation time $\tau$($T$,$E$) vs $T$ and $E$.}
\end{figure}

As discussed below,  our main focus is thermoelectric transport along the $a$ axis, which is by far the most efficient. Since spontaneous polarization points along the $c$ axis, any field from that source will not affect transport along $a$ by symmetry. Analogously, the matrix element of piezoelectric electro-phonon scattering, which is proportional to the $a$ component of dielectric displacement \cite{ridley1},  is also zero by symmetry \cite{nye,electrom,anewFE}.

\begin{figure}[ht]
\centerline{\includegraphics[width=0.8\linewidth]{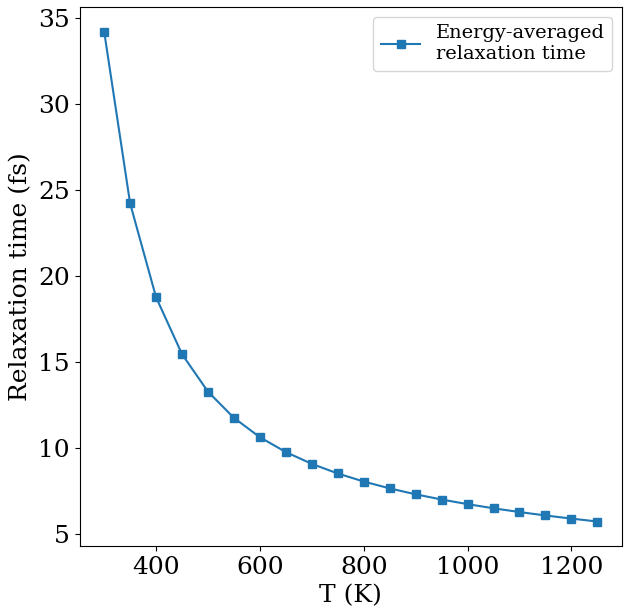}}
\caption{\label{taufig2} Energy-averaged scattering time $\tau$($T$) vs $T$.}
\end{figure}

Of course, $ZT$ will now depend on $T$ via all the ingredients in Eqs.\,\ref{ZTT} to \ref{scrat}, and on materials parameters appearing in the scattering rates. 
These parameters are imported from experiment or previous calculations or calculated directly. For the former group, high-frequency \cite{epsinf}  and lattice \cite{iniguez} dielectric constants $\varepsilon_{\infty}$=5  and the $\varepsilon_{\rm lattice}^{ii}$=(62, 44, 65) tensor diagonal; average sound velocity  $v$=5.2 km/s \cite{soundv}; density 5870 kg/m$^3$ \cite{matproj}; for the latter group, effective conduction mass $m_c^*$=0.2 $m_e$ and valence mass $m_v^*$$\simeq$$m_e$, dominant LO-phonon  frequency $\hbar$$\omega_{\rm LO}$=58 meV, for which see Ref.\cite{discLO}.  The behavior of $\tau$($E$,$T$)  is sketched vs $E$ and $T$ in Figure \ref{taufig1} { (the raw data are provided for further perusal in the Supplementary Material)}. The averaged time $\tau_{\rm ave}$($T$) is in Figure \ref{taufig2}. To compare with  constant relaxation-time results, we simply recalculate the various quantities with a constant time which we arbitrarily choose to be $\tau_0$=$\tau_{\rm ave}$($T$=300 K), i.e. equal to the energy-average time at room T.

The use of an energy-averaged time is equivalent by construction to assuming $\tau$ vs $E$ to be a constant (a different one for each $T$); since typical electron energies are in the low-energy region (below, say, 30 to 40 meV from the band edge) where $\tau$($T$,$E$) is large, our averaging over all energies will tend to underestimate $\tau$ and therefore  give conservative (smaller) estimates of $ZT$.
 This procedure is a vast improvement over the constant-time approximation. It is, however, of inferior quality (though much simpler) than calculations such as those in Ref.\cite{noi2} which include the  model energy-dependent relaxation time in the Onsager integrals over energy \cite{noi2,bt2} (and of course, inferior to ab initio electron-phonon coupling calculations of $\tau$ \cite{epw}). 
 
 In particular, in this approximation, $S$ does not depend on $\tau$, which is not generally the case in reality; this should not be too serious an issue given the specific shape of $\tau$ as function  of energy. For example, in Mg$_3$Sb$_2$ in the degenerate regime \cite{noi2}, $S$ has the same $T$ dependence for constant and ($E$,$T$)-dependent $\tau$, except from a practically rigid shift. Again, if Mg$_3$Sb$_2$ results are  any indication, $S$ (and hence $ZT$) may be underestimated in our approach.

\subsection{Lattice thermal conductivity}
As indicated by Eq.\ref{ZTT}, an essential ingredient  { in the calculation $ZT$ is the lattice thermal conductivity $\kappa_{\ell}$. Below we fix $\kappa_{\ell}$$\simeq$1.2 W/(K m) and $T$-independent, as observed experimentally in Ref.\cite{klto}.
To justify this choice, and  make our theoretical description more self-contained, we investigate the reasons for the small and weakly T-dependent $\kappa_{\ell}$}. Given the near-impossibility of a direct calculation for this large-unit-cell material, we estimate it from the ab initio  $\kappa_{\ell}$ of BaTiO$_3$  in a geometrically constrained configuration -- that is, we calculate $\kappa_{\ell}$(T) for the bulk and for a film that is 1.5 nm thick, i.e. as thick as the individual perovskite block in LTO.  
To this end, beside anharmonic and isotope-disorder scattering, a Casimir boundary scattering rate is included \cite{tk,ziman} in the form
$$
P_{{\bf q}j,{\hat {\bf n}}}^{\rm C}=\frac{c_{{\bf q}j}\cdot\hat{\bf n}}{FL} \, \overline{n}_{{\bf q}j}
(\overline{n}_{{\bf q}j}+1),
$$
with $\overline{n}$ the equilibrium phonon occupations, $L$ the film thickness, $F$ a correction factor between 0.5 and 1 (we set it to 1), $c_{{\bf q}j}$ the group velocity of  mode $j$ at wavevector {\bf q}, and  $\hat{\bf n}$ the normal to  the film. { (The factor $F$ is set to unity; often it is assumed to be 0.5, which would amount to assume a thickness of 3 nm).} The anisotropy of the material is thus included projecting  each phonon velocity along the 'short' direction: phonons traveling orthogonal to the plane of the film are scattered the most, and those in the plane of the film not at all; importantly, all intermediate directions of phonon wavevector are scattered proportionally to the projection, so most end up being affected.

{ The physical rationale for this model is that {\it i)} Ba is mass-wise the closest element to La in the periodic table (excluding rare earths), and {\it ii)} the local structure of both BaTiO$_3$ and LTO are made up of TiO octahedra, and therefore most likely share the main vibrational features. LTO is characterized in addition by the internal interfaces, which are modeled by the Casimir term. }

 { We do not consider the possible occurrence of mass disorder (a potential source of thermal conductivity reduction: see e.g. the discussion in Ref.\cite{sige} or \cite{noi1}), as it does not appear to be significant in the high-quality material of Ref.\cite{klto}. Also,  similar $\kappa_{\ell}$ values have been recorded \cite{soundv} in the LTO-isostructural layered perovskite Sr$_2$Nb$_2$O$_7$, reinforcing the case for a geometric cause to the small $\kappa_{\ell}$. }

\begin{figure}[ht]
\centerline{\includegraphics[width=0.8\linewidth]{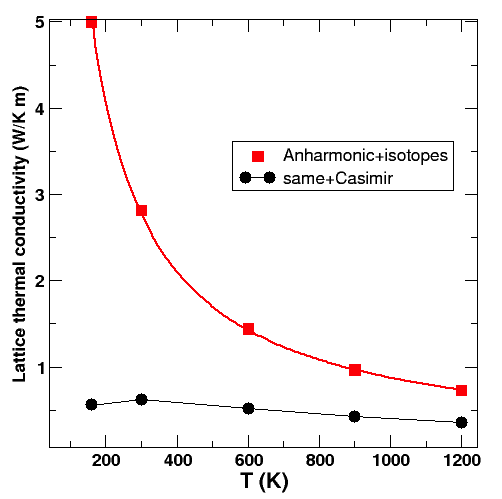}}
\caption{\label{tk} Lattice thermal conductivity vs $T$ for  BaTiO$_3$ in bulk and film form.}
\end{figure}

To calculate $\kappa_{\ell}$ for BaTiO$_3$, similarly to e.g. Ref.\cite{noi1}, we employ the {\tt Quantum Espresso} suite \cite{QE} and two of its extensions \cite{tk,D3Q},   optimized norm-conserving pseudopotentials from the PseudoDojo site \cite{pseudodojo}, the 
generalized gradient approximation \cite{pbe} to density functional theory, and a plane-wave basis  cutoff  of  100 Ry.  We use the  energy-force code to obtain the relaxed configuration in the ground state rhombohedral phase, with a (8$\times$8$\times$8) k-points grid; the  phonon code to obtain the dynamical matrix on the same grid; the {\tt D3Q} \cite{D3Q} code to obtain the third-order anharmonic force constants on a commensurate (4$\times$4$\times$4) grid; and the {\tt TK} code \cite{tk} for the phonon Boltzmann equation using a (12$\times$12$\times$12) grid and $\delta$-functions width of 5 cm$^{-1}$. The isotope abundance is taken from the NIST database \cite{nist}. For simplicity, BaTiO$_3$ is assumed to have the low-T rhombohedral structure throughout.

 Figure \ref{tk} shows the inverse average of the  $\kappa_{\ell}$ tensor for bulk BaTiO$_3$ and for film-like BaTiO$_3$ with thickness 1.5 nm (i.e. the thickness of a perovskite block in LTO). The finite size  causes a sharp decrease in $\kappa_{\ell}$, and largely suppresses its $T$-dependence. Importantly, the $\kappa_{\ell}$ tensor is essentially isotropic, i.e. the size reduction is effective even in the plane of the film.
Since our calculation is  underestimating the BaTiO$_3$ bulk $\kappa_{\ell}$ (of order 5 W/(K m) at room T \cite{batio3}) by 30-40\%, our prediction for the film should be corrected to $\kappa_{\ell}$$\simeq$1 W/(K m). This is in fair agreement with experiment, and supports the  idea that the low lattice thermal conductivity  of LTO and analogous compounds (Sr$_2$Nb$_2$O$_7$ behaves quite similarly) is due to the  effective confinement of phonons  within the stacked blocks of the layered structure.
{ As mentioned we adopt henceforth a constant $\kappa_{\ell}$$\simeq$1.2 W/(K m). If we used the theoretical $\kappa_{\ell}$, the changes would be minor;   
ZT could  be slightly higher at high temperature, due to the slight decrease 
of $\kappa_{\ell}$ with $T$.} 

\section{Results}
\subsection{Dependence on doping and temperature of ZT and other tensors}	

We first discuss the various quantities making up $ZT$, and $ZT$ itself, as function of carrier density. Here we concentrate on $n$-doping  (briefly touching upon $p$-doping in Sec.\ref{secmiscell2}) and on the $a$ component of the various tensors, since the other components (see below) turn out to produce a much smaller, hence uninteresting, $ZT$.


\begin{figure}[ht]
\includegraphics[width=0.49\linewidth]{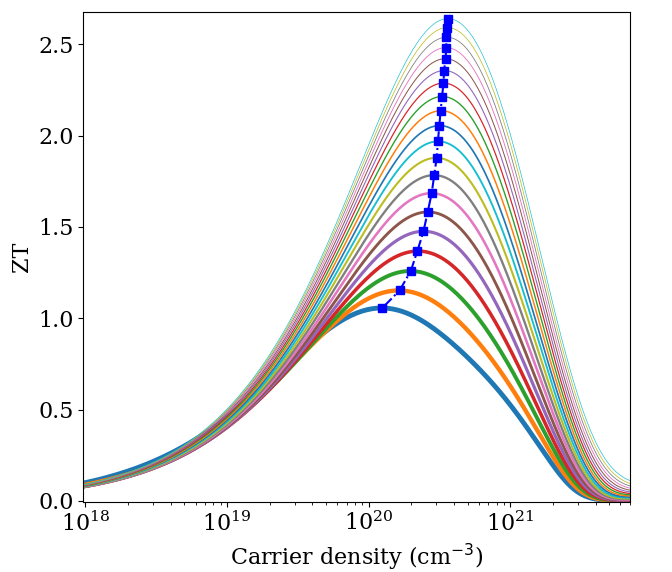}
\includegraphics[width=0.49\linewidth]{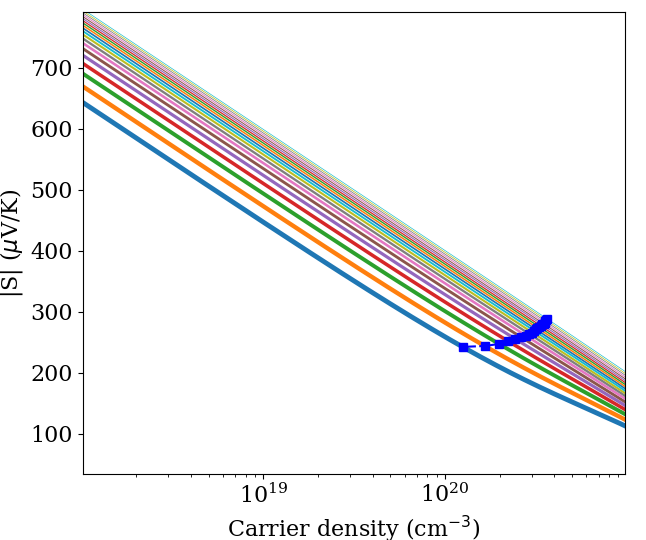}\\
\includegraphics[width=0.49\linewidth]{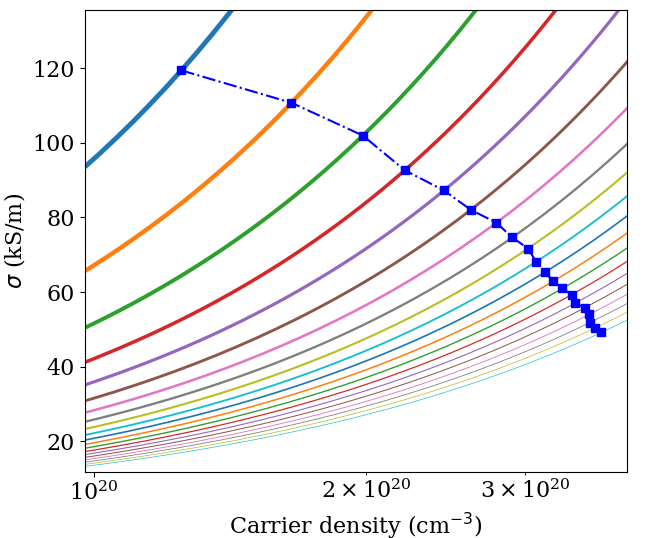}
\includegraphics[width=0.49\linewidth]{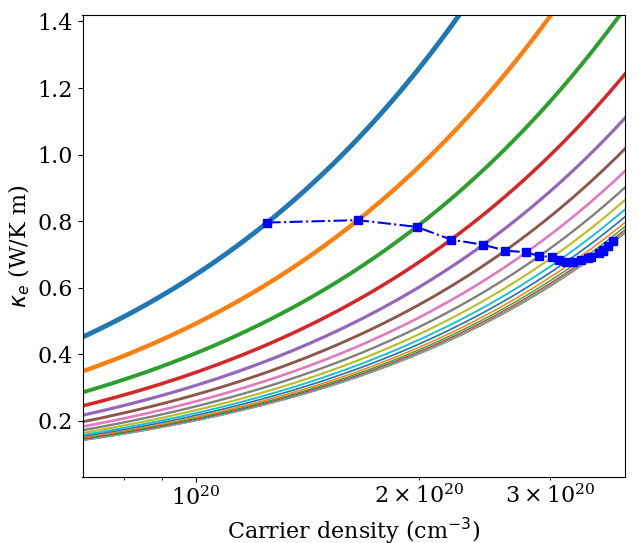}
\caption{\label{allvdop} The $a$ component of (top left to bottom right) the $ZT$, Seebeck, electrical conductivity, and thermal conductivity  tensors vs $n$-type carrier density for $T$ between 300 to 1250 K (thick to thin lines) in steps of 50 K. Squares indicate the value at optimal doping (i.e. the one maximizing $ZT$).}
\end{figure}

The results are summarized in Figure \ref{allvdop}. All four quantities ($ZT$, $S$ in absolute value, $\sigma$, $\kappa_e$) are displayed as function  of excess $n$-type carrier density, and for a set of temperatures between 300 and 1250 K, as solid lines. Different $T$'s are represented by line thickness, which  decreases as $T$ increases in steps of 50 K. For each temperature, squares connected by dash-dotted lines indicate the value of each quantity at the optimal doping, i.e. that at which $ZT$ is a maximum. The maximum temperature is chosen so as to avoid approaching too closely the ferroelectric transition of LTO (about 1500 K).

In the top left panel, $ZT$ is seen to be between over 1 and about 2.6 in the chosen temperature range (see also below). This obviously interesting $ZT$ results, as usual, from a combination of factors: a good Seebeck (top right) of about 250-300 $\mu$V/K, an almost $T$-independent total thermal conductivity (the electronic component is  bottom right in the Figure) of less than 2 W/(K m), and a significant conductivity (bottom left) of roughly 40 to 100 kS/m (i.e. a resistivity about 10 to 25 $\mu$$\Omega$$\cdot$m).  This happens despite the strong decrease in relaxation time with $T$ (indeed, $ZT$ increases severalfold if a constant time is used, see Sec.\ref{secmiscrelt}). { We recall that the conductivities are determined solely by the crystal bands, and by scattering from phonons (mostly polar) and charged impurities as dictated by $\tau$($T$); no other scattering mechanism (such as atomic disorder, dislocations, neutral impurities and traps, etc.) is accounted for.}

The optimal carrier density (or, familiarly, doping) is by definition the density at which the maxima of $ZT$  occur. Optimal doping is clearly between 1 and
3$\times$10$^{20}$ cm$^{-3}$; of course it remains to be seen if such optimal doping  densities can be achieved experimentally.  In that regard, it is interesting to note that the Seebeck coefficient can achieve much  larger values at lower doping, in particular over 650 $\mu$V/K at 10$^{18}$ cm$^{-3}$ at  room $T$ (and increasing with $T$). This is clearly interesting for applications requiring just a large $S$, as this relatively low doping should be fairly easy to achieve. { We point out that the $T$-dependent values of $ZT$ and related quanties at different, non-optimal densities  can be easily read off Fig. \ref{allvdop}; for example at 10$^{20}$ cm$^{-3}$, $ZT$ is between 1 and 2, and 
at 10$^{19}$ cm$^{-3}$ it is 0.4 to 0.5.}


\begin{figure}[ht]
\includegraphics[width=0.49\linewidth]{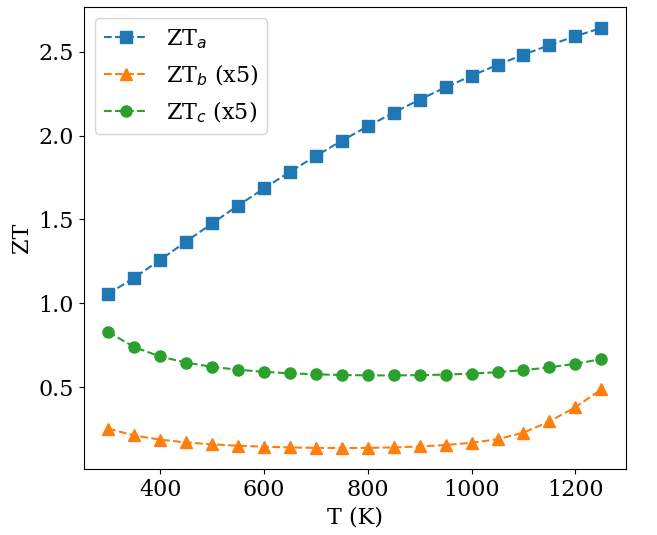}
\includegraphics[width=0.49\linewidth]{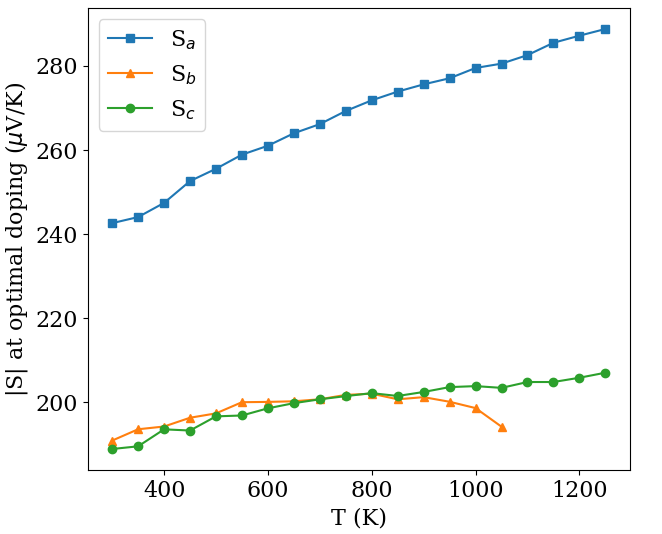}\\
\includegraphics[width=0.49\linewidth]{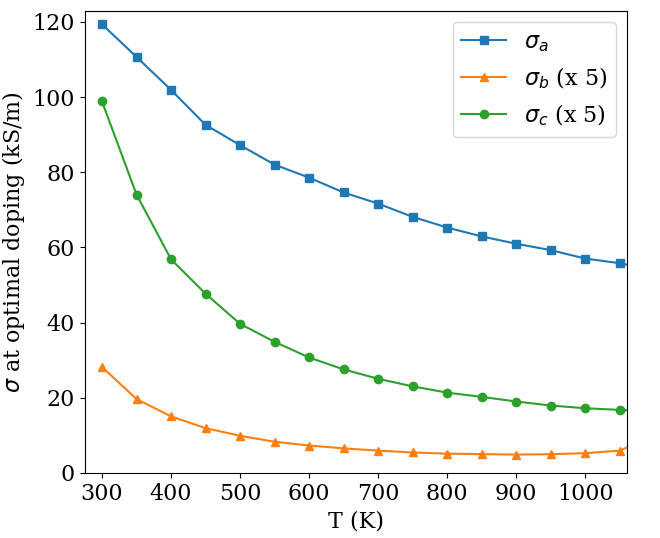}
\includegraphics[width=0.49\linewidth]{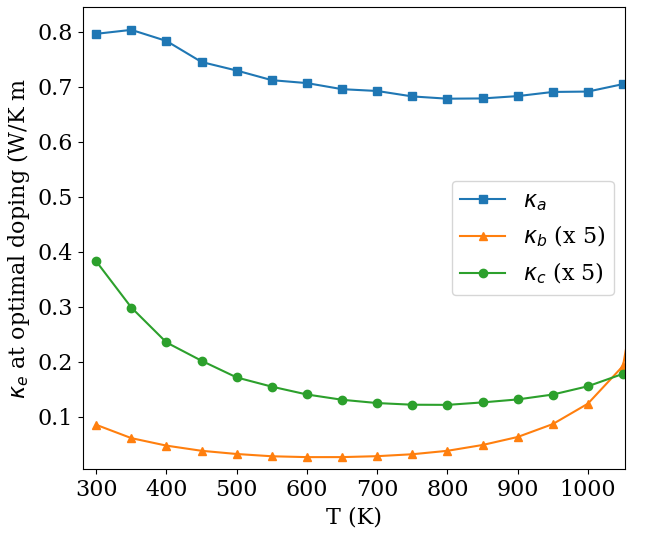}
\caption{\label{allvT} Components of the $ZT$, Seebeck (absolute value), electrical conductivity, and electronic  thermal conductivity tensors vs. $T$ at optimal $n$-doping for each $T$.}
\end{figure}

We now show directly  the temperature dependence of the various quantities in Figure \ref{allvT}. These are the values corresponding to optimal doping, i.e. basically those overlayed in Figure \ref{allvdop} as squares. Here all components are considered, to give a feel for their relative size. For the Seebeck coefficient, the $a$ component discussed previously is over 20\% larger than the other two, and all tend to increase with $T$. For the other quantities, the $a$ component exceeds the other two by at least an order of magnitude, or more. The conductivity is suppressed by the decrease in relaxation time, as is the thermal conductivity, though to a much lesser extent. This combination results in the $b$ and $c$ components of $ZT$ being flatter in T, or even decreasing slightly, besides being much smaller in absolute value. 
In experiments on sintered ceramic LTO (Ref.\cite{expseeb}, Fig.8) a $|S|$  between roughly 160 and 270 $\mu$V/K was observed in the 300-800 K range; in the same range, the trace of our Seebeck tensor is between 180 to 220  $\mu$V/K, which can be considered fairly good agreement given the different materials conditions in the two cases.

\subsection{Component average}

Thermoelectrics are often fabricated in polycrystalline form, especially with a view to reducing lattice thermal   conductivity. On the other hand, LTO is pretty robustly crystalline in its usual growth conditions; also, the thermal conductivity is low in the crystal already (as we discussed previously) due to the strongly layered structure. 

\begin{figure}[ht]
\centerline{\includegraphics[width=0.8\linewidth]{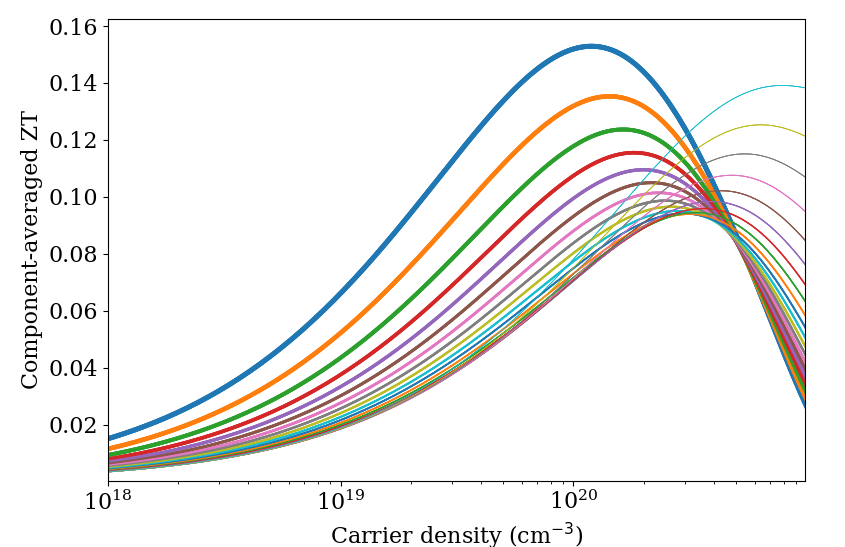}}
\caption{\label{ZTave} $ZT$  calculated from the inverse average of  $\sigma$ and $\kappa$ tensors, and the trace of the Seebeck tensor (display conventions as in Fig.\protect\ref{allvdop}).}
\end{figure} 

However, an estimate of the average of the various quantities is still of interest due to the strong anisotropy.
In Figure \ref{ZTave} we plot vs carrier density and temperature, with the same conventions as in Figure \ref{allvdop}, an average $ZT$  calculated with the usual expression, but using the Mathiessen inverse average of the electrical and thermal conductivities and the trace of the Seebeck coefficient (the lattice thermal conductivity was assumed constant and isotropic as found in experiment as well as calculations). Clearly, the maximum direction-averaged $ZT$ is not  especially interesting.  

\subsection{Constant relaxation time}
\label{secmiscrelt}%

\begin{figure}[ht]
\centerline{\includegraphics[width=0.8\linewidth]{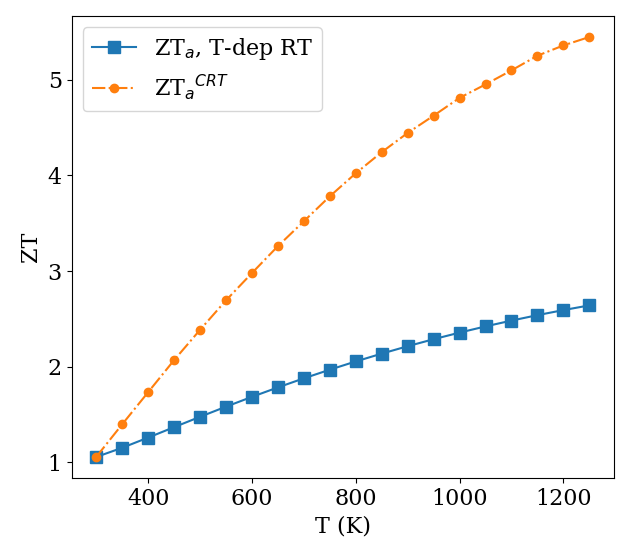}}
\caption{\label{ZTvCRT} The $a$ component of $ZT$ calculated with $T$-dependent and constant relaxation time.}
\end{figure}

To demonstrate the significance of the $T$ dependence of the relaxation time, in Figure \ref{ZTvCRT} we compare $ZT$ calculated with the $T$-dependent relaxation time and with a constant relaxation time, which we choose to be the value of the $T$-dependent $\tau$ at 300 K, i.e. $\tau_0$=34 fs. Since $\tau$($T$) decreases monotonically with $T$, the constant-time   $ZT$ is much larger at all $T$s, reaching almost 5.5 at the end of range. 

\subsection{{\it p}-doping}
\label{secmiscell2}
\begin{figure}[ht]
\includegraphics[width=0.49\linewidth]{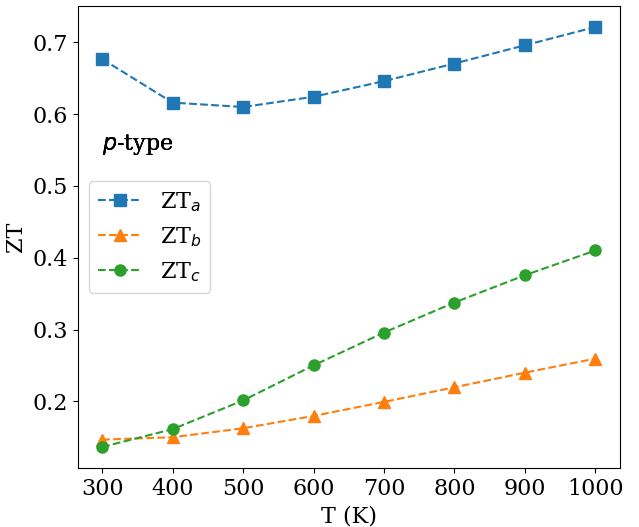}
\includegraphics[width=0.49\linewidth]{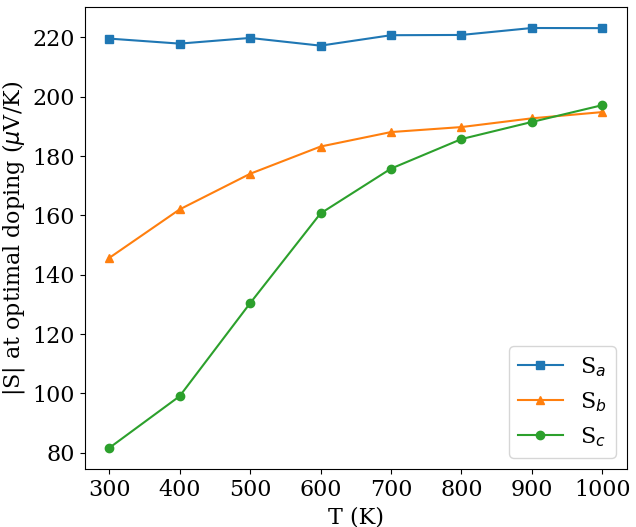}
\caption{\label{allvTp} Components of $ZT$ and Seebeck 
vs. $T$ at optimal $p$-doping for each $T$.}
\end{figure}

The potential of LTO in the $p$-type case is less remarkable than in the $n$-type case. For one thing, $ZT$ only reaches about 0.7 in the best instance, as seen in Figure \ref{allvTp}; for another thing, the optimal doping is higher for $p$-type: 
  the $a$ component requires carrier densities in the high 10$^{20}$ cm$^{-3}$ 
range, and well into the mid 10$^{21}$ cm$^{-3}$ for the other components.


	The Seebeck coefficient is somewhat smaller than in $n$-type at optimal doping, but again that is due to the high optimal density; as in $n$-type, $S$ is high at low density: 700 to 800 $\mu$V/K, depending on $T$, at 10$^{18}$ cm$^{-3}$, up to nearly 1 mV/K at 10$^{17}$ cm$^{-3}$. Again, as in the $n$ case, the practical feasibility of $p$-doping remains to be ascertained.


%
 
\section{Summary}

We have predicted a  thermoelectric figure of merit between 1 and 2.5 in the $T$ range of 300 to 1200 K  under $n$-doping in the layered perovskite La$_2$Ti$_2$O$_7$ via  calculations of the electronic structure, transport coefficients, and thermal conductivity.  The optimal carrier density is in the low-10$^{20}$ cm$^{-3}$ range. At that density the Seebeck thermopower coefficient is between 200 and 300 $\mu$V/K; it can, however, reach  nearly 1 mV/K at lower doping. The largest $ZT$ is obtaned along the $a$ crystal axis, while the other components are one to two orders of magnitude smaller. The maximum $ZT$ in $p$-type conditions is a factor of 2 to 4 smaller than in $n$-type, and it requires carrier densities about an order of magnitude higher. Much of the potential of this material is due to its small and almost $T$-independent   lattice thermal conductivity; using a model based on ab initio anharmonicity calculations,  we explain this low value as due  to effective phonon confinement within the   layered-structure blocks.

\section*{Acknowledgments}
Work supported in part by Universit\`a di Cagliari, Fondazione di Sardegna, Regione Autonoma Sardegna via  Progetto biennale di ateneo 2016 {\it Multiphysics approach to thermoelectricity}, and CINECA-ISCRA grants.


\begin{thebibliography}{99}
\bibitem{discLO}
We obtain the zone-center TO frequencies via a frozen-phonon   VASP calculation, and thence the LO frequencies by the Lyddane-Sachs-Teller relation with the $T$=0 theoretical \cite{iniguez}  lattice dielectric constants. We include in the scattering rate only the lowest LO mode with polarization along the $a$ axis, which affects transport in the $a$ direction. Phonons with polarization along $b$ or $c$ are slightly lower in energy, and will reduce further the $b$ and $c$ components of $ZT$, which are 
uninterestingly small anyway. We do not account for the $T$-dependence of the phonon frequencies, or for the increase of static dielectric constant increases with $T$ due to the incipient ferroelectric transition (occurring at about 1500 K). Our LO frequency, therefore, is probably underestimated; accordingly, the relaxation time is overestimated, and $ZT_a$ is underestimated.
\end{thebibliography}
\end{document}